\documentclass[preprint2]{aastex}

\shorttitle{High Resolution Spectroscopy of H {\sc II} Galaxies}
\shortauthors{Telles, Mu\~noz-Tu\~n\'on \& Tenorio-Tagle}

\newcommand{\simlt}{\mathrel{\spose{\lower 3pt\hbox{$\mathchar"218$}}
     \raise 2.0pt\hbox{$\mathchar"13C$}}}
\newcommand{\simgt}{\mathrel{\spose{\lower 3pt\hbox{$\mathchar"218$}}
     \raise 2.0pt\hbox{$\mathchar"13E$}}}
\newcommand{\casiana}{Mu\~noz-Tu\~n\'on}
\newcommand{\kms}{\thinspace\hbox{$\hbox{km}\thinspace\hbox{s}^{-1}$ }}
\newcommand{\sqcm}{\thinspace\hbox{$\hbox{cm}^{2}$}}
\newcommand{\ergs}{\thinspace\hbox{$\hbox{erg}\thinspace\hbox{s}^{-1}$}}
\newcommand{\ergsqcmsec}{\thinspace\hbox{erg}\sqcm\thinspace\hbox{s}$^{-1}$}

\newcommand{\ha}{\hbox{$\hbox{H}\alpha$ }}
\newcommand{\hb}{\hbox{$\hbox{H}\beta$ }}

\newcommand{\etal}{{\sl et\nobreak\ al.\ }}

\newcommand{\Mpc}{\thinspace\hbox{Mpc}}
\newcommand{\kmsec}{\thinspace\hbox{$\hbox{km}\thinspace\hbox{s}^{-1}$}}
\newcommand{\kmsecmeg}{\thinspace\kmsec\Mpc$^{-1}$}
\newcommand{\aandas}{Astr.\ Astro\-phys.\nobreak\ Suppl.\nobreak\ }

\newcommand{\apspsc}{Astro\-phys.\nobreak\ Sp.\nobreak\ Sc.\nobreak\ }

\received{}
\revised{}
\accepted{}
\slugcomment{}

\begin{document}

\title{High resolution spectroscopy of H {\sc II} Galaxies:\\
Structure and Supersonic line widths}

\author{Eduardo Telles}
\affil{Observat\'orio Nacional,
                Rua Jos\'e Cristino, 77,
                20921-400 - Rio de Janeiro -
                BRASIL}
\email{etelles@on.br}
\author{Casiana Mu\~noz-Tu\~n\'on}
\affil{Instituto de Astrof\'{\i}isica de Canarias, E-38200 La Laguna, Tenerife, SPAIN}
\email{cmt@ll.iac.es}
\and
\author{Guillermo Tenorio-Tagle}
\affil{INAOE, Apartado Postal 51, Puebla, Pue. MEXICO}
\email{gtt@inaoep.mx}

\begin{abstract} 

We present high resolution echelle spectroscopy of a sample of H {\sc II}
galaxies.  In all galaxies we identify different H$\alpha$ emitting
knots along the slit crossing the nucleus. All of these have been
isolated and separately analyzed through luminosity and size {\it vs}
$\sigma$ diagnosis plots. We find that in all cases, for a particular
galaxy, the bulk of emission comes from their main knot and therefore,
at least for the compact class galaxies, we are dealing with,
luminosity and sigma values measured using single aperture
observations would provide similar results to what is obtained with
spatially resolved spectroscopy. In the size {\it vs} $\sigma$ plots
as expected there is a shift in the correlations depending on whether
we are including all emission in a single point or we split it in its
different emitting knots. The problem of a proper determination of the
size of the emitting region so that it can be used to determine the
mass of the system remains open. From the data set gathered, using the
highest surface brightness points as recently proposed by
Fuentes-Masip et al. (2000), the best luminosity {\it vs} $\sigma$
correlation turns out to be consistent with a Virial model.

\end{abstract}

\keywords{
- galaxies: starburst -- galaxies: structure --
galaxies: line width
}

\section{Introduction}

H {\sc II} galaxies are a subsample of dwarf galaxies with a strong
star formation activity leading to intense nebular lines easy to
detect in objective prism surveys.  The optical properties of H {\sc
II} galaxies are dominated by their strong emission line spectra
superposed on a weak blue continuum. The question, posed by Sargent \&
Searle (1970) of whether these are young galaxies forming stars for
the first time seems now to have been answered by recent studies.  H
{\sc II} galaxies show an underlying stellar population of
intermediate to old ages (Telles \& Terlevich 1997, Doublier \etal
1997, Marlowe \etal 1999, Cair\'os \etal 2000). The optical/near-IR
colors of their underlying galaxies are similar to late type dwarf
galaxies, although their structural parameters derived from brightness
profiles indicate that H {\sc II} galaxies are more compact than late
type quiescent dwarfs (Telles \& Sampson 2000; Telles \etal 1999).

Their morphological classification splits them into two groups: Type
I includes the most luminous H {\sc II} galaxies, all of which present
an overall irregular morphology. Types II's, on the other hand, are
more compact and rounder and present a luminosity profile similar to
that of dwarf spheroidal galaxies (Telles, 1995).  The few H {\sc II}
galaxies found to have a bright neighbor (maybe by chance) are all 
Type II's of regular morphology, contrary to what one would expect if
galaxy interactions could be held responsible for the morphological
disturbances found in Type I's (Telles \& Terlevich 1995).  { In
addition, H {\sc II} galaxies as a class are not preferentially clustered
around faint low mass galaxies (Telles \& Maddox 2000), but some are
found to have HI companions (Taylor
1997). The triggering mechanism of the present starburst in H {\sc II}
galaxies is not at all clear, yet for these dwarfs internal processes
must play a major role.}

Here we show echelle data obtained at the William Herschell telescope
(4.2m) at the "Observatorio del Roque de los Muchachos" (ORM) at La
Palma on a sample of type II dwarfs.  Our high resolution echelle data
show a strong variation in the line profiles across the emitting
region and even in the most compact sources, there is an indication of
separate knots of star formation (SF) evolving concurrently within the
galaxy nucleus.  { The presence of multiple knots of star formation
within the line emitting regions are also observed on high spatial
resolution images, in particular in the near-IR.  Telles \etal (1999, 2000)
also identified knots which are possible super stellar clusters (SSC).
Their individual properties may impose further constraints on the
history of SF of a galaxy as well as indicate possible mechanisms able to
trigger the present burst.}

For our sample here we have defined different apertures across the H$\alpha$
luminosity profile and have extracted the spectra corresponding to
different emitting knots in every object. Also the {\it total} spectra
was obtained to mimic single aperture observations (Melnick \etal
1988), when it includes the ensemble of individual knots.

The structure found in H {\sc II} galaxies has profound implications
on several topics. In particular on issues such as star formation and
its possible sequential propagation in H {\sc II} galaxies, and how is
the ISM structured in these galaxies. Another issue central in this
field of research is the validity of the interpretation, and use of
the empirical correlations, of size and luminosity {\it vs} their
supersonic line widths for high redshift galaxies.  These correlations
were first found for Giant H {\sc II} Regions by Melnick (1979),
Terlevich \& Melnick (1981) and Melnick \etal (1987) and extended to H
{\sc II} galaxies by Melnick \etal (1988) and Telles \& Terlevich
(1993).  The fact that Giant H {\sc II} Regions and H {\sc II}
galaxies have supersonic motions and follow the same correlations have
made us use and refer to the whole emsemble of objects as a single
family.  This may indicate that process of massive cluster formation
is similar in both H {\sc II} galaxies and Giant H {\sc II} Regions
and it is the main reason as to why we will not differenciate between
them in discussing their dynamical properties.  Here we describe in
Section~\ref{data} and ~\ref{analysis} our selected sample and
observational strategy, and Section~\ref{results} presents full detail
of the results for each galaxy. A general discussion and a summary of
the results is given on Section~\ref{conclusion}.

\section{The Observations}\label{data}

We have carried out long-slit echelle observations of a sample of 7
H {\sc II} galaxies (see Table 1) with the Utrecht Echelle spectrograph at
the 4.2-m William Herschel Telescope at the ``Roque de los Muchachos''
observatory, La Palma, Spain, on the nights of 30-31/05/1997.
Exposures of 1200s were obtained using a 2148$\times$2148 Tek CCD and
the 79.0 lines/mm echelle, reaching typical of S/N per pixel $\ge $ 30 at
H$\alpha$.  The orders covered are 55 to 30, from $\sim$ 4130 \AA~ -
7340 \AA, including \hb and \ha with short slit 19$''$ mode (Mrk 36,
UM 461, UM 533, Mrk 930) and order 34, from $\sim$ 6500 \AA\ - 6650
\AA\ on \ha with long slit 250 $''$ mode (Mrk 59, II Zw 40, VII Zw
403).  The spectral resolution (FWHM) of our setup was 0.13 \AA~in the \hb
region and 0.18 \AA\ in the \ha region. This led to a resolution of
$\sim$ 8 \kms.  The CCD frames were reduced using standard procedures
with the tasks in IRAF.  The data was debiased, trimmed and flat-field
using a normalized flat field produced with apflatten.  The spectra
were extracted using doecslit with appropriate parameters.  Wavelength
calibration was performed by using the comparison spectrum of
Thorium-Argon arcs. Flux calibration was accomplished through the
observation of standard stars.  The atmospheric extinction correction
was also applied at the time of calibration using the mean extinction
curve for La Palma.  The seeing was less than 1.5 arc seconds.

Table 1 lists the global properties of our selected
galaxies. Columns 1 gives the galaxy name. Columns 2 and 3 give their
coordinates. Columns 4-7 present their spectroscopic information from
the literature: redshift, H$\beta$ flux, equivalent width of H$\beta$,
and the Balmer decrement (Terlevich etal 1991, SCHG).

\begin{deluxetable}{ccccccc}
\tabletypesize{\footnotesize} 
\tablecaption{Spectroscopic data from
Terlevich etal 1991, SCHG. \label{tab1}}
\tablewidth{0pt}
\tablehead{
\colhead{\bf name} & \colhead{\bf R.A.} & \colhead{\bf $\delta$} &
\colhead{\bf z} & \colhead{\bf $-lgF(H\beta)$} & \colhead{\bf
$W(H\beta)$} & \colhead{\bf $C(H\beta)$} \\
\colhead{}   & \colhead{(1950)} & \colhead{(1950)}  &\colhead{}  & 
\colhead{[\ergsqcmsec]} & \colhead{[\AA]} & \colhead{}     
}
\startdata
 Mrk 59 &  12 56 38.2 & 35 06 53 & 0.0027 & 14.26 & 17  &0.54\\ 
 IIZw70  & 14 48 55.2 & 35 46 36 & 0.0040 &       &     &0.40\\ 
 VIIZw403& 11 24 35.8 & 79 16 03 &-0.0003 &       &     &0.40\\ 
 Mrk 36  & 11 02 15.6 & 29 24 31 & 0.0022 & 13.36 &  70 &0.44\\ 
 UM 461 & 11 48 59.4 &--2 05 41  & 0.0035 & 13.20 & 342 &0.40\\ 
 UM 533  & 12 57 25.0 &  2 19 08 & 0.0030 & 13.59 & 101 &0.76\\ 
 Mrk930  & 23 29 29.3 & 28 40 16 & 0.0181 & 13.05 & 71  &0.55\\ 
\enddata							     
\end{deluxetable}

\section{Data analysis}\label{analysis}

\begin{figure*}
\figurenum{1}
\epsscale{2.0}

\bigskip

\centerline{\sc Figure 1 =  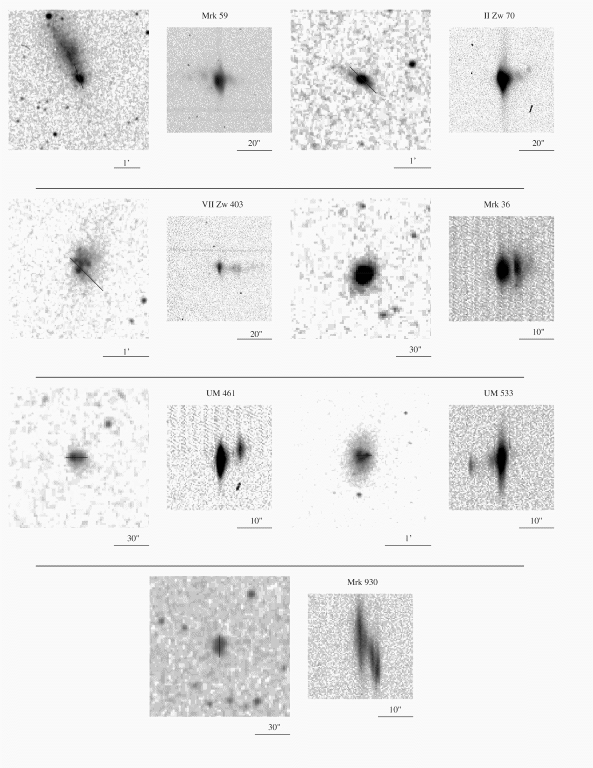}


\caption{Left larger panel: Direct image from
DSS (North is up, East is left) with slit location indicated.  Right
smaller panel: Echelle CCD image on the H$\alpha$ spectral region
(horizontal is the spatial direction, vertical is the dispersion
direction). \label{fig1}}
\end{figure*}

Figure~1 displays, for each galaxy, two panels.  The larger left panel
shows the direct image from DSS of a galaxy, while the right smaller
panel (alongside the direct image) shows its Echelle CCD 2D spectrum.
Superimposed on the direct image the location and position angle of
the slit for our observation is indicated.  In all cases the slit was
centered on the brightest nucleus, as defined by broad band
images. For those galaxies with a compact appearance the slit was
aligned E-W, and for those which displayed a more extended structure,
the slit was placed along the ``extended'' orientation.  In the
Echelle images, the horizontal direction represent the spatial
dimension while the vertical is the dispersion direction.  The image
scales are given for both panels.  {The precise location in relation
with the peak intensity is given in Table~\ref{tab2}.  Note that the
spectrum of every subregion has been extracted using variable size
apertures covering the extent of the H$\alpha$ emission.  In all cases
a large aperture including all the emission detected along the slit in
each object has also been defined and subsequently referred to as the
{\it total} emission.}

The spectral range covered on each spectra includes H$\alpha$
($\lambda$6563\AA) and [NII]$\lambda\lambda$6583, 6548\AA\ lines. A
clear H$\alpha$ emission is present on the spectra from all regions on
all galaxies; The [NII] doublet however only appears on some regions
of three galaxies (Mrk59, IIZw70 and UM533). Thus, the analysis
presented here is based on the behavior of the H$\alpha$ emission line
of each galaxy.

There is a wide variety of spectral patterns. Some of them display a
very well-behaved Gaussian H$\alpha$ line profile; others however,
present clear signs of multiple spectral components. These, have been
fitted using two (three in the case of IIZw70) Gaussian templates.

Table~\ref{tab2} lists the derived properties of the spectra of each
of the recognized knots or regions as well as the {\it total} for each
galaxy. Column 1 gives the galaxy name. Column 2 gives the {\it
spatial} nomenclature. A clear peak in emission covered by the
aperture is referred to as a "knot", while "region" could represent a
zone between knots or an extension. Total emission is the aperture
covering all the emission arising from a galaxy.  Column 3 gives the
rotator position angle of the observation which corresponds to the
slit position angle shown in Figure~1.  Columns 4-5 provide the
aperture position in relation to the main knot and size in
arcsecs. Columns 6-8 refer to the analysis of the spectra and give the
parameters obtained from Gaussian fits to the H$\alpha$ emission line,
namely: H$\alpha$ flux, central wavelength and line width ({\sc FWHM}
in \AA).  The typical error in flux are 10-20\% estimated from the
sensitivity curves from the standard star observations and $\sim$
5-10\% in line width from the comparison with the H$\beta$ line in the
echelle observations or externally with different frames when
available. In the cases where more than one Gaussian is needed to
account for the H$\alpha$ emission line, the Table provides the
various parameters derived for each of the components. The last column
in Table~\ref{tab2} indicates the {\it spectral} nomenclature showing
in some cases the presence of the additional spectral components from
the emission line profile and their relative importance. A featureless
emission line well fitted by a single Gaussian is referred as
\,``single\,'', while \,``global\,'' is a single fit to an emission
line that shows more than one resolved component. This would mimic an
observation with poorer spectral resolution.  The main and
secondary components of an emission line were also fitted whenever
they arose.

\begin{figure}
\figurenum{2}
\epsscale{1.0}
\plotone{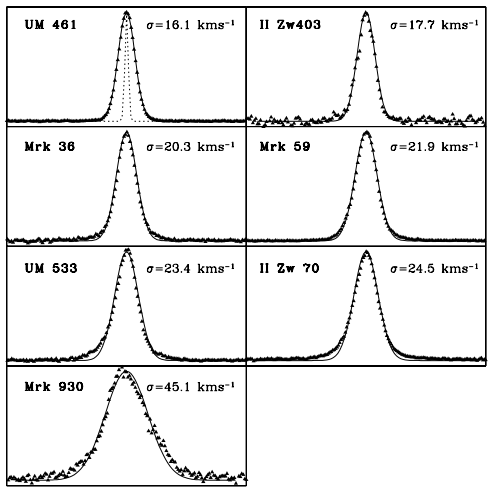}
\caption{Emission line profiles  of HII galaxies 
covering a range 10 \AA~ centered on the H$\alpha$ line. Also shown
are the Gaussian fits (solid lines) used to derive the final gas
velocity widths which are given in the upper-right side of each panel
and listed in Table~\ref{tab3}. The comparison Thorium lamp line
profile is shown in the first upper-left panel as a dotted line.
\label{fig2}}
\end{figure}

The {\it global} (or {\it single}) Gaussian fit is a very good
representation of the {\it total} H$\alpha$ line profile in spectrum
of a galaxy, as shown in Figure~\ref{fig2}. The results of the fit can
be compared with parameters traditionally used to analyze the
kinematics of H {\sc II} regions by using single aperture spectroscopy
(see Melnick \etal 1988 and \casiana~ 1994 and references therein).

\section{Results}\label{results}

\begin{deluxetable}{lccrrcccc}
\tabletypesize{\scriptsize}
\tablecaption{Derived spectral properties. \label{tab2}}
\tablewidth{0pt}
\tablehead{
\colhead{\bf name} & \colhead{KNOT ID} & \colhead{PA} & \multicolumn{2}{c}{Aperture [$``$]} &
\colhead{F(H$\alpha$)} & \colhead{Peak $\lambda$} & \colhead{FWHM} & \colhead{\bf Spectral}\\
\colhead{} & \colhead{} & \colhead{} & \colhead{pos.} & \colhead{size} & \colhead{[\ergsqcmsec]} & 
\colhead{[\AA]} & \colhead{[\AA]}  &  \colhead{\bf components}  \\ 
}
\startdata
{\bf Mrk59} &{\it total}& 20 & &30.6 & 3.356E-13  & 6580.7 & 1.190 &   single \\ 
            &           &    & &     & 9.270E-15  & 6601.4 & 1.170 &  single[NII] \\
              &Main knot& &0.0 & 10.6 & 3.710E-13 & 6580.7 & 1.173 & single \\ 	
              & region 2& &-17.6& 6.2 & 3.270E-15 & 6581.1 & 0.870 & single \\ 		 
              & region 3& &-8.2 & 5.2 & 1.730E-14 & 6581.2 & 1.347 & {\it global}	  \\	 
              &         & &     &     & 1.720E-14 & 6581.2 & 1.347 & Main component\\  
              &         & &     &     & 2.860E-15 & 6579.8 & 0.800 & Secondary component \\ 
              & region 5& & 6.1 & 5.0 & 1.670E-14 & 6580.8 & 1.018 & {\it global}\\
              &         & &     &     & 1.590E-14 & 6580.8 & 0.960 & Main component \\        
              &         & &     &     & 2.540E-15 & 6579.6 & 1.360 & Secondary component \\					  
\hline							  
{\bf IIZw70} &{\it total}& 45& & 19.9 & 2.686E-13 & 6589.2 & 1.320 &   single \\
     &         &           &   &      & 1.550E-14 & 6610.0 & 1.800 & single[NII]\\
             &Main knot&   &0.0& 11.3 & 2.667E-13 & 6589.2 & 1.313 & {\it global}\\ 
     &         &           &   &      & 1.680E-13 & 6589.2 & 0.850 & Main component\\              
     &         &           &   &      & 4.510E-14 & 6589.9 & 1.120 & Secondary component  \\             
     &         &           &   &      & 6.000E-14 & 6588.4 & 1.020 & Secondary component.  \\            
           & region 2  & &10.18& 8.6  & 5.374E-14 & 6589.2 & 1.423 &  {\it global}\\
     &         &         &    &      & 4.360E-14 & 6589.3 & 1.180 &  Main component\\   
     &         &         &    &      & 4.060E-15 & 6590.5 & 1.480 &  Secondary component  \\
     &         &         &    &      & 9.870E-15 & 6588.3 & 1.640 &  Secondary component  \\
            & region 3&  &5.14&7.5   & 1.695E-14 & 6589.5 & 1.437 & single\\
\hline
{\bf VIIZw403}&{\it total} & 45 &    &18.4 & 5.650E-14 & 6560.7 & 0.980  & single \\
                & Main knot&    &0.0 & 9.7 & 2.754E-14 & 6560.7 & 0.986  & single \\      
                  & knot 2 &    &7.8 & 3.7 & 8.938E-15 & 6560.6 & 0.829  & single \\	   
                  & knot 3 &    &11.1& 3.7 & 4.560E-15 & 6560.5 & 1.048  & {\it global} \\
                  &        &    &    &     & 3.960E-15 & 6560.5 & 0.900  & Main component \\ 
                  &        &    &    &     & 4.290E-16 & 6559.6 & 0.610  & Secondary component \\           
                 &extension&    &20.7&23.9 & 7.968E-14 & 6560.7 & 1.013  & poor S/N \\
\hline
{\bf Mrk36} &{\it total} & 90 & &15.1 & 1.650E-13 & 6577.4 & 1.108 & {\it global} \\ 
            &      &    &  &     & 1.100E-13 & 6577.4 & 0.910 & Main component \\	 
            &      &    &  &     & 5.840E-14 & 6577.4 & 1.870 & low-intensity wings\\ 
         & Main knot&      &0.0&  8.1& 1.115E-13 & 6577.4 & 1.036 & {\it global}\\                 
            &      &       &   &     & 9.040E-14 & 6577.4 & 0.920 & Main component \\
            &      &       &   &     & 2.640E-14 & 6577.4 & 2.270 &  low-intensity wings\\
                & knot 2&  &4.3&  4.0& 4.721E-14 & 6577.5 & 1.336 &  single \\
            & extension&   &7.9&  5.0& 1.129E-14 & 6577.6 & 1.373 &  {\it global}\\ 
\hline
{\bf UM461} &{\it total}&90 & &  11.4 & 2.085E-13 & 6586.1 & 0.906 &  single \\        
        & Main knot &  &0.0 &6.1 & 1.938E-13 & 6586.1 & 0.905 &  single \\        
        & knot 2 &     &5.6 &4.4 & 1.449E-14 & 6586.8 & 0.870 &  single \\        
\hline
{\bf UM533} &{\it total}& & 90 &14.8  & 1.309E-13 & 6582.8 & 1.263 &        {\it global} \\
            &           &  &   &      & 1.190E-13 & 6582.8 & 1.130 &  Main component \\     
        &       &          &   &      & 1.350E-14 & 6581.6 & 1.180 &  low-intensity wings \\ 
        &       &          &   &      & 7.260E-15 & 6603.5 & 0.960 &  [NII]\\		 
        &Main knot&        &0.0& 7.4  & 1.146E-13 & 6582.7 & 1.258 &  single\\	 
        &       &          &   &      & 6.740E-15 & 6603.5 & 1.110 &  [NII]\\
            & knot 2 &   &-8.7 &3.0   & 6.822E-15 & 6582.6 & 0.979 &  single \\
       & interknots&     &-4.5 &4.4   & 5.927E-15 & 6582.3 & 0.789 &  single \\
\hline
{\bf Mrk930}&{\it total} &180& &   13.4 & 1.508E-13  & 6682.1 & 2.432  &  single\\      
         & Main knot & &  0.0   &    5.9 & 8.609E-14 & 6682.4 & 2.433  &  single\\	          
         & knot 2    & &  1.2   &    1.7 & 2.670E-14 & 6681.9 & 1.703  & single\\	          
         & knot 3    & &  3.2   &    2.0 & 4.197E-14 & 6681.1 & 1.679  & single\\	          
         & knot 4    & &  4.8   &    2.0 & 3.150E-14 & 6680.7 & 1.494  & {\it global} \\
         &           & &     &        & 2.316E-14 & 6680.5 & 1.160  & Main component\\         
         &           & &     &        & 8.574E-15 & 6681.5 & 1.460  &  Secondary component\\

\enddata 
\end{deluxetable}

Here we briefly describe the most general spectral features found in
each galaxy of the sample. The qualitative description of each of
these features is based on the precise measurements ie, the emission
line parameters, given in Table~\ref{tab2}.

From our results in Table~\ref{tab2} it is clear that even
the apparently most compact of our H {\sc II} galaxies (e.g. UM461; which
spans across 11") splits into several components, or emitting knots.
Some of these may be true centers of stellar formation but others
could simply be high density condensations ionized from the outside
by the stellar radiation.

The emitting knots in every galaxy present different properties
(intensity, line width, etc.)  however in all cases the velocity
dispersion extracted from the $total$ fit to the spectra of every
galaxy is, within the errors, identical to the velocity dispersion
inferred for the main emitting knot found in every galaxy.  This is a
result now well established in the field of giant H {\sc II} regions
(e.g. Mu\~noz-Tu\~non 1994, Sabalisck \etal 1995, Mu\~noz-Tu\~non \etal
1995).  In the seven galaxies in our sample the main knot peak
intensity amounts to half, or more, of the total intensity derived for
the total fit.  The other main components of other knots, in
every galaxy, amount to less than an order of magnitude of the peak
intensity of the total fit, and in many cases their line width is
smaller than that derived for the main emitting knot.  This is also
the case for all secondary components required to fit low intensity
line asymmetries, or shoulders, although the peak intensity of these
lines is usually much smaller.  On the other hand, the secondary
components required to fit the wing emission of some of the knots in
every galaxy contribute by much less than an order of magnitude to the
total emission and are in all cases very broad lines with line width
values well above the value measured for the main emitting knot. Note
however that uncertainties in the fit to low intensity wings are much
larger.

Thus, the line emitting properties of H {\sc II} galaxies, including
their supersonic $\sigma$ value are in all our cases dominated by the
emitting properties of the most intense knot. And thus, similarly to
giant H {\sc II} regions, it seems that the main emitting knot
reflects an intrinsic property generated by the recent starburst event
{(see Tenorio-Tagle \etal 1993, and Mu\~noz-Tu\~non \etal 1996)}.

\subsection{Notes on Individual Objects}

\subsubsection{Mrk 59}

Mrk 59 has a NE-SW oriented "comet-like" optical appearance (see
Figure~1). More recent studies have played special
attention to this sub-class of H {\sc II} galaxies (see Noeske \etal
2000). It shows a conspicuous bright region, which clearly dominates
the emission, located at the tip of an extended tail. The optical
nucleus contains a rich substructure in nebular emission.  We have
defined five sub-regions in total. A clear peak in luminosity (main
knot) is however identified with a size (measured on the baseline of
the H$\alpha$ line spatial profile) of $\sim$ 10 arcseconds. At a much
lower intensity level, there is an extended emission centered $\sim$ 7
arcseconds from the main knot. In a similar way we have identified
other intensity peaks showing emission along the slit which we named
 regions 2 and 3.  In this galaxy the main knot intensity totally
dominates the output of the emitting region.

\subsubsection{II Zw 70}

This is a very compact dwarf. Its image displays a single optical knot
with some "disk-like" extended emission in broad band (see
Figure~1). On the long slit Echelle CCD frame, one can
see a clearly defined H$\alpha$ line as well as the [NII] doublet.
The nebular emission spatial distribution is almost point-like with a
slight asymmetry presenting an extended, lower intensity $H_\alpha$
emission. The nebular luminosity profile, led us to define three
apertures from which we have extracted and analyzed the spectra. We
refer to them as main knot (which overwhelmingly dominates the emission), 
region 2 and region 3.  The H$\alpha$
emission line in the three apertures presents a remarkably symmetric
profile, which however cannot be well fitted by a single Gaussian due
to the presence of low-intensity wings. In fact a minimum of three
Gaussians are needed to obtain a good fit. The values derived for the
three components are given in Table~\ref{tab2}.

\subsubsection{VII Zw 403}

This is a classical example of a very nearby blue compact galaxy. 
Its broad band optical morphology presents a typical
core-halo structure with an extended light distribution.  It has been
the target of HST for studies of its stellar content (Schulte-Ladbeck
\etal 1999a, 1999b).  The distance to VII Zw 403 derived from these
studies, and adopted here, is 4.4 Mpc.

From our data, we have identified along the slit a clear peak in
emission: the main knot; which produces half of the luminosity of the 
galaxy.  The spectrum of the main knot  is well fitted by a single
Gaussian with the possible presence of low intensity wings, too weak
for being modeled. No signs of line splitting are seen at our present
spectral resolution. Besides the main knot there are two additional
intensity maxima (knots 2 and 3) very close to each other. The
spectrum from knot 2 displays a Gaussian-like $H\alpha$ line. On knot
3 however, there seems to be a secondary component. The line can be
reproduced by using two Gaussians, very close in wavelength (less than
1\AA). Moving along the slit, immediately behind the two secondary
knots one finds an extended emission (named ``extension'' in
Table~\ref{tab2}), which despite the poor S/N we have fitted
using a Gaussian, with parameters similar to those obtained from the
fit to the {\it total} spectrum of the galaxy.


\subsubsection{Mrk 36}

Our echelle observations were taken with the slit oriented E-W (P.A.=
90) crossing the optical core of the galaxy. This object has been of
particular interest because it shows a very ``blobby'' structure, most
visible in near-IR images (Telles \etal 2000).  However, our present
observation crosses the bright nucleus which, as described below, also
presents a well resolved structure.  Bi-dimensional spectroscopy with
Integral Field Units will be very valuable in the near future to
unveil its internal kinematics and help us answer the question of the
origin of the present starburst.

From the luminosity profile across the slit we have identified two
main peaks of nebular emission (see Figure~1), labeled
as main knot (the brightest one) and knot 2. Two more apertures have
been defined for the analysis; {\it total} includes the
whole emission from the galaxy and an extension which
rescues the extended, low intensity emission besides knot 2.

The main knot, shows the dominant H$\alpha$ emission component.  For
the fitting to be successful also at the base of the line
(low-intensity wings), a second Gaussian is needed. This has a much
lower intensity but it is broader than the main component. The
parameters obtained from doing the fitting with only one Gaussian are
also presented on Table 2 and referred to as {\it global}. The secondary,
low intensity, broad component is also present in the integrated ({\it
total}) spectrum, which is clearly dominated by the emission from the
brightest main knot.

\subsubsection{UM 461}

UM461 appears along the slit as a point-like source (see
Figure~1).  We extracted its spectra from an aperture
covering the main knot. A secondary much weaker second peak, west from
Main knot was also detected (knot 2). To compare the line profiles with
those obtained using a single aperture spectrum, a third aperture
covering all emission have been taken ({\it total}). The H$\alpha$
emission from the three apertures can be nicely fitted with a single
Gaussian profile.  The main knot dominates the total nebular emission of
the galaxy and therefore the spectra from the main knot and the total are
rather similar. It is important to notice that the second-weak knot
(knot 2), although clearly spatially separated from the main one,
shares its main kinematic parameters (see Table~\ref{tab2}).

\subsubsection{UM 533}

Its optical appearance, as seen at the DSS is that of a halo which
engulfs a heart-like core with two clearly visible nuclei. Our slit
was positioned in the E-W direction (P.A.=90) crossing the two
nuclei. First thing to notice (see Figure~1) is that
these two peaks are also visible in nebular emission. There the
contrast is much larger, with a clear bright knot (Main knot) and a
much weaker one, separated by about 10 arcsec (knot 2). To make a
comparison and to find the spatial trend of the emission lines two
more apertures were defined: an aperture covering the region in
between knots 1 and 2 (interknots) and an aperture including the whole
emitting region ({\it total}). In this galaxy [NII] (6583 \AA) is also
detected on the bright knot. The H$\alpha$ emission of the brightest
knot shows a good Gaussian behavior on the line core, however at least
a second Gaussian is needed to account for the emission arising from
the line base. Low intensity wings are also evident on the spectra
covering the whole emission, which, as mentioned before, is dominated
by the brightest knot.

\subsubsection{Mrk 930}

This galaxy presents the largest redshift ($z$ = 0.018) of our sample.
It has a knotty appearance on the broad band image, and its nebular
emission along the slit (located N-S, P.A.=180) also displays a
multiple knot structure. The ensemble of knots seem to show at the
same time a velocity trend (see Figure~1). From the
luminosity profile four apertures, centered on each of the emission
peaks (knots), were selected. The spectra from all of them are well
reproduced with a single Gaussian profile, whose parameters are given
in Table 2. The {\it total} spectrum, with a spatial beam of 14
arcseconds, covering all emission is also Gaussian-like, and presents the
same $\sigma$ value detected in the main knot. The main knot also accounts 
in this case 
for more than half of the total emission from the galaxy.

\subsection{Structural parameters}

Our echelle data has
led us to built a comprehensive data base of sizes, velocity
dispersion and luminosities of the galaxies of our sample; not only of
the total parameters but also for the starbursts
sub-structures found on each of them  (Table~\ref{tab3}).

Table~\ref{tab3} describes the derived physical parameters of our
sample.  Columns 1-2 identify the galaxy and the region or knot as
described in Table~\ref{tab2}.  Column 3 gives values of the velocity
dispersion $\sigma_{gas}$ obtained after correcting the observed line
width (given in Table~\ref{tab2}) for instrumental and thermal
broadening ($\sigma^2_{gas} = \sigma^2_{obs} - \sigma^2_{inst} -
\sigma^2_{th}$), assuming a 10$^4$ K gas.  Column 4 \& 5 give the
H$\alpha$ luminosity and the physical ``radius'' (Aperture size / 2)
of a region or knot in pc.  Column 6 gives an estimate of the
H$\alpha$ surface brightness within the slit.  Finally we give derived
distances from their observed redshift $z$ (expect for VII Zw 403 as
described above).  Throughout this paper we use the current value of
H$_0 = 65$\kmsecmeg (Suntzeff \etal 1999).

Our high quality data is here confronted to see if it is in agreement
with previous statistical works which show that H {\sc II} galaxies
follow a luminosity {\it vs} $\sigma$ ($L$ $\propto$ $\sigma^4$) and
the size {\it vs} $\sigma$ ($R \propto$ $\sigma^{2}$) relationships.
These correlations have often been obtained using single aperture
spectroscopy and their validity under better spectral and spatial
resolution is a key issue to be verified.  Note that in each of our
galaxies, the difference in the velocity centroids of the various
components is in all cases less than 1 \AA ~(from Table~\ref{tab2}),
and thus, the velocity shift between components cannot account on its
own for the velocity dispersion detected in the total spectrum of a
galaxy.  The velocity dispersion in our data set thus reflects an
intrinsic property of each of the components and in some cases this is
indeed supersonic ($\sigma_{gas}$ $>$ c$_{\rm H {\sc II}}$; the sound
speed in the ionized gas).

\begin{deluxetable}{lcccccc}
\tabletypesize{\footnotesize}
\tablecaption{Physical Parameters derived from the {\it single} Gaussian
fits to the H$\alpha$ line (see text for details). We use
H$_0$=65\kmsecmeg. \label{tab3}}
\tablewidth{0pt}
\tablehead{
\colhead{\bf name} & \colhead{Knot ID} & \colhead{$\sigma_{gas}$}&
\colhead{log(L(H$\alpha$))}& \colhead{Radius} & \colhead{log SB} &
\colhead{Distance} \\
\colhead{} & \colhead{}& \colhead{[km s$^{-1}]$} & \colhead{[erg
s$^{-1}$]} & \colhead{[pc]} & \colhead{[erg s$^{-1}$ pc$^{-2}$]} &
\colhead{[Mpc]} 
}
\startdata
Mrk 59    &total     &  21.96&     40.16&    923&   34.95&  12.4 \\
          &Main knot &  21.62&     40.20&    318&   35.45&  12.4 \\
          &knot 2    &  15.37&     38.17&    190&   33.63&   \\
          &knot 3    &  25.11&     38.90&    163&   34.43&   \\
          &knot 5    &  18.46&     38.86&    153&   34.43&   \\
II Zw 70  &total     &  24.51&     40.31&    889&   34.81&  18.4 \\
          &Main knot &  24.37&     40.31&    507&   35.05&   \\
          &region2   &  26.56&     39.61&    387&   34.47&   \\
          &region3   &  26.84&     39.12&    339&   34.03&   \\
VII Zw 403&total     &  17.76&     38.39&    196&   34.30&   4.4 \\
          &Main knot &  17.88&     38.08&    104&   34.26&    \\
          &knot2     &  14.58&     37.59&     39&   34.19&    \\
          &knot3     &  19.16&     37.30&     39&   33.90&    \\
          &extension &  18.44&     38.54&    255&   34.33&    \\
Mrk 36    &total     &  20.32&     39.61&    372&   34.88&  10.2 \\
          &Main knot &  18.84&     39.44&    198&   34.98&   \\
          &knot2     &  24.92&     39.07&    100&   34.91&   \\
          &extension &  25.66&     38.46&    126&   34.19&   \\
UM 461    &total     &  16.10&     40.09&    450&   35.07&  16.2 \\
          &Main knot &  16.08&     40.06&    238&   35.32&   \\
          &knot2     &  15.34&     38.96&    177&   34.33&   \\
UM 533    &total     &  23.41&     40.00&    498&   35.00&  13.9 \\
          &Main knot &  23.31&     39.94&    248&   35.25&   \\
          &knot2     &  17.64&     38.71&    101&   34.41&   \\
          &interknot &  13.64&     38.64&    144&   34.19&   \\
Mrk 930   &total     &  45.06&     41.48&   2729&   34.95&  83.8 \\
          &Main knot &  45.08&     41.23&   1197&   35.07&   \\
          &knot2     &  31.21&     40.72&    341&   35.10&   \\
          &knot3     &  30.76&     40.91&    407&   35.22&   \\
          &knot4     &  27.21&     40.79&    405&   35.10&   \\ 
\enddata
\end{deluxetable}

\subsubsection{The [$L - \sigma$] relation}

\begin{figure}
\figurenum{3}
\epsscale{1.0}
\plotone{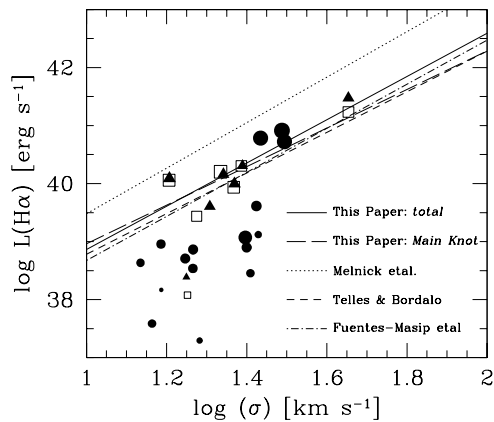}
\caption{The luminosity - line width relation. Points:
{\it total} apertures (solid triangles), Main Knot (open squares),
other regions or knots (solid circles).  Point sizes represent
relative H$\alpha$ surface brightness. The lines are simple linear
square fits for our data (total: solid line, Main knot: long dashes)
and the statistical works from the literature. \label{fig3}}
\end{figure}

Figure~\ref{fig3} plots our L(H$\alpha$) {\it vs} $\sigma$ for
{\it all} the data from Table~\ref{tab3}.  The points were
derived from single Gaussian fits to the spectral line of each
aperture named {\it global} or {\it single} in the last column of
Table~\ref{tab2}.  Measurements inferred from {\it total}
apertures are shown as solid triangles. Open squares represent the
values of the Main knot in each galaxy, while solid circles are other
regions or secondary knots.  In the plot, the size of the points is
proportional to their H$\alpha$ surface brightness, (see
Table~\ref{tab3}).

Clearly the luminosity of the points representing the main knots (open
squares) falls very close to values derived from apertures covering
the {\it total} extent of the emitting region (solid triangles).  The
implication of this is that the brightest knot in every galaxy
dominates almost entirely the total luminosity. In addition, the
velocity dispersion inferred for the main knots is almost identical to
that derived from the total galactic emission.

A second thing to notice is the clear correlation displayed by either
the total emitting spectra or by the main knots in our sample of
galaxies (see Figure 2 and Table 4).  This fit holds for all galaxies
in our sample with the exception of VII Zw 403 which presents a
log$\sigma \sim 1.25$ and a logL(H$\alpha$) $\sim$ 38.  Based on the
work of Fuentes-Masip \etal (2000), where a thorough analysis of the
luminosity and size {\it vs} velocity dispersion of the giant H {\sc
II} regions in the large irregular galaxy NGC 4449, led them to
realize that the correlations only hold for nebulae with a supersonic
line width and a surface brightness above 2 $\times 10^{35}$ erg
s$^{-1}$ pc$^{-2}$, we excluded this galaxy from our analysis. Note
that VII Zw 403 is the galaxy with the lowest surface brightness in
the sample.  For this reason, we conclude that this object must fall
below the threshold for which the relations hold (although the true
luminosity threshold value cannot be established here). The surface
brightness effect could be related to a second parameter in the
relations, again similar to the fundamental plane of elliptical
galaxies as Telles \& Terlevich (1993) first attempted to investigate.

As far as the correlation is concerned (lines in
Figure~\ref{fig3}), we note that our present data either {\it
total} (solid line) or {\it main knot} (long-dashed line) closely
agree with the results from other works.  Fuentes-Masip \etal (2000)
results are shown as dot-dashed line.  Telles \& Bordalo (2000)
analyzed a sample of about 40 H {\sc II} galaxies and their
preliminary results are shown as short-dashed lines.  The similarity
of these fits is evident in Figure~\ref{fig3}.  In the light of
our results, we conclude that irrespectively of the structure that a
galaxy (or a giant H {\sc II} region) may have, the main emitting knot
is likely to be sitting at the bottom of the gravitational potential
well of its host galaxy, and therefore the overall motions are
dominated by the mass of the complexes of gas and stars.

The existence of these relations and the presence of a surface
brightness effect prompt us to suggest that the L {\it vs} $\sigma$
relation may have an associated second parameter.  This issue is being
further investigated using 2D spectroscopy.

 The numerical results of the various fits are shown in
 Table~\ref{tab4}.  Note however that the slope of the original
 Melnick \etal (1988) (L {\it vs} $\sigma$) relation is virtually
 identical to all the others, although there is a clear offset on the
 zero point of their correlation, after all appropriate
 transformations are applied. This could be due to a systematic error
 in the calibration of their data, since independent calibration of
 Telles \& Bordalo (2000) for a sample of a statistically significant
 sample of H {\sc II} galaxies and the redshift independent zero
 points given by  Fuentes-Masip et al. (2000) and Bosch (1999)  for
 Giant H {\sc II} Regions all agree within the uncertainties with
the zero point measured here.

\begin{deluxetable}{lccc}
\tabletypesize{\footnotesize}
\tablecaption{The [$L - \sigma$] simple least square to the total
aperture and global line fit values.  Also shown are the relations
found from the literature using the same simple linear least square
fit and all transformed to H$_0$=65\kmsecmeg as used throughout this
paper. \label{tab4}}
\tablewidth{0pt}
\tablehead{
\colhead{}  & \multicolumn{3}{c}{$L \propto x \times \log(\sigma)$} \\
\colhead{} & \colhead{$x$}  &  \colhead{$\Xi$}[\ergs] & \colhead{\rm RMS} 
}
\startdata
{This paper}& 3.73$\pm$1.01 &  35.14$\pm$1.40 & 0.275   \\
Melnick \etal & 3.92$\pm$0.47 & 35.55$\pm$0.67 & 0.426 \\
Telles \& Bordalo & 3.50$\pm$0.67 & 35.28$\pm$1.03 & 0.539 \\
Fuentes-Masip \etal & 3.80$\pm$1.20 & 34.90$\pm$1.70 & \\ 
\enddata
\end{deluxetable}

\subsubsection{The [$R - \sigma$] relation}

\begin{figure}
\figurenum{4}
\epsscale{1.0}
\plotone{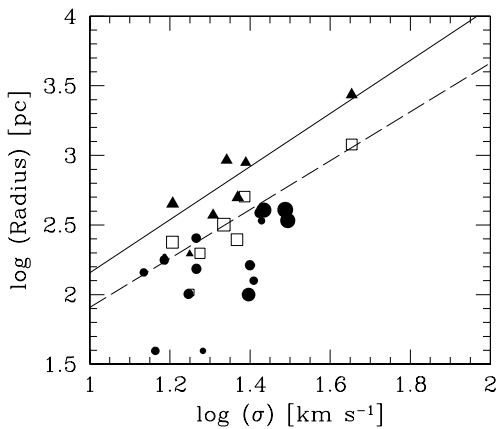}
\caption{The Size - line width relation. Points: {\it total}
apertures (solid triangles), Main Knot (open squares), other regions
or knots (solid circles).  Point sizes represent relative H$\alpha$
surface brightness. The lines are simple linear square fits for our
data (total: solid line, Main knot: long dashes). \label{fig4}}
\end{figure}

Figure~\ref{fig4} shows the (R {\it vs} $\sigma$) relation for our
present sample of H {\sc II} galaxies.  Again, measurements from {\it
total} apertures are shown as solid triangles. Open squares represent
the values for the main knots, while solid circles are other regions
or knots.  The sizes of the points are proportional to their
corresponding H$\alpha$ surface brightness as given in
Table~\ref{tab3}.

From the analysis of this plot and the peak $\lambda$ of all regions
from Table~\ref{tab2} we conclude that no ordered motions
are detected in these galaxies.  The measured line widths are unlikely
to be due to rotation, because as one goes down to smaller sizes
(apertures) one still measures the same $\sigma$ value (but see Van
Zee \etal, 1998, who reach a different conclusion.)

As mentioned in many previous works about the topic, it is difficult
to define a radial scale.  Here we are considering the aperture size
for each extraction of a spectrum which must be a scaled value to the
true size of the emitting region. Even though, it is noticeable from
Figure~\ref{fig4} that there is a clear correlation between size
and the velocity dispersion for these objects, again, points with low
surface brightness seem to fall out of the correlations.  Also, there
is a clear similarity between the slopes of the fit to the main knots
(open squares, long-dashed line) and the total galaxy points (solid
triangles, solid line).  Both are very close to what one would expect
for virialized systems. The results of the fits are given in
Table~\ref{tab5}.  In addition, we note that these results
(R$_{total}$ and R$_{Main Knot}$ {\it vs} $\sigma$) are compatible
with the findings of Telles (1995) for the size {\it vs} $\sigma$
relation for H {\sc II} galaxies, namely that the Effective Radius
(derived from surface photometry) {\it vs} $\sigma$ relation is a
scaled version of the ``Core'' Radius (from luminosity profiles) {\it
vs} $\sigma$ relation in a sample of about 40 galaxies, while the
latter present a smaller scatter.

\begin{deluxetable}{lccc}
\tabletypesize{\footnotesize} 
\tablecaption{The [$R - \sigma$] simple least square to the total and Main
Knot apertures vs. global line fit values.\label{tab5}}
\tablewidth{0pt}
\tablehead{
\colhead{}  & \multicolumn{3}{c}{$R \propto y \times \log(\sigma)$} \\
\colhead{} & \colhead{$y$}  &  \colhead{$\Phi$}[pc] &\colhead{\rm RMS}  
}
\startdata
{\it Total}& 1.90$\pm$0.46 &  0.25$\pm$0.63 & 0.126   \\
{\it Main Knot}& 1.75$\pm$0.35 & 0.15$\pm$0.49 & 0.090 \\
\enddata
\end{deluxetable}

\section{Discussion}\label{conclusion}

Without a doubt, one of the most intricate issues in the field of
giant H {\sc II} regions, and now also of H {\sc II} galaxies, is the
origin of their supersonic velocity dispersion ($\sigma_{gas}$ $>$
$c_{\rm H {\sc II}}$).

It has long been established that a simple agglomeration of a large
number of H {\sc II} regions cannot lead to the dominant supersonic
line profiles, and thus, giant H {\sc II} regions, and H {\sc II}
galaxies, constitute a different class of objects. Different not only
because of the size of their emitting regions and the fact that they
are powered by recent violent bursts of star formation, but also
because of their essence: their peculiar inner dynamics.

It is now well established that the supersonic line width correlates
with the size of the emitting region (size $\propto$ $\sigma^{2}$) and
luminosity ($L$ $\propto$ $\sigma^4$) of the ionized regions
(Terlevich \& Melnick 1981, Melnick \etal 1987,1988, Telles \&
Terlevich 1993). The fact that the correlations are similar to the
relations inherent to virialized stellar systems such as globular
clusters, spiral bulges and the cores of elliptical galaxies, led
Terlevich \& Melnick to postulate that giant H {\sc II} regions and H
{\sc II} galaxies are themselves virialized systems and thus that the
measured gas velocity dispersion ($\sigma_{gas}$) should directly
relate to their total mass. In their original scenario Terlevich \&
Melnick envisaged a gravitational potential that forced the collective
motion of clumps of gas to present the supersonic $\sigma$
values. However, in our sample and in the recent finding in giant H
{\sc II} regions the massive emitting knots present a relative
velocity unable to explain the observed supersonic $\sigma$ values
(see Tenorio-Tagle \etal 1993 1996, and references therein).  Instead,
the brightest knots present an intrinsic value of sigma formerly
ascribed to the whole ionized region.  This fine tuning regarding the
size of the supersonic $\sigma$ region and its luminosity has brought
the empirical correlations into a much better agreement with what is
expected from a virialized system.  On the theory ground,
virialization is now believed to have lots to do with the motion of
the low-mass stars moving in the gravitational potential of the system
while undergoing winds. This enhances their cross-section and allows
them to cause the stirring of the gas left-over from the star
formation event (see Tenorio-Tagle \etal 1993).

On the other hand, some authors have claimed that the supersonic line
widths could result from a plethora of unresolved expanding shells
caused by the mechanical energy of the large number of massive stars
powering each of the sources (see Chu \& Kennicutt 1994 and references
therein). In the latter case however, there is no obvious way to
explain the empirical correlations and thus it has been argued that
the measured line widths follow the correlations simply due to a lack
of resolution in the integrated spectra obtained from single aperture
observations (but see also Tenorio-Tagle \etal 1996).

Our results confirm and extend the empirical correlations found for
giant H {\sc II} regions and H {\sc II} galaxies.  This fact has
profound implications both on the observations of H {\sc II} galaxies
and on the interpretation of their supersonic line width.  Enhanced
spectral and spatial resolution seems to unveil an intricate structure
in H {\sc II} galaxies.  Note that the previously measured (single
aperture) supersonic motions in fact arise from regions of much
smaller dimension than that occupied by the full extent of the ionized
gas.  As shown in Figure~\ref{fig3} and Figure~\ref{fig4}
for compact H {\sc II} galaxies, accurate determinations of the size
and luminosity of the region presenting the supersonic $\sigma$ values
nicely outline the correlations. H {\sc II} galaxies when resolved,
present several emitting knots with a variety of shapes, luminosity
and $\sigma$ values.  However, we have shown that in these cases the
global integrated value agree very closely with the properties derived
for the main emitting knot.  This is simply because the intrinsic
properties (luminosity, velocity dispersion) of a galaxy are dominated
by the central (core) component.

A fine calibration of these relations for local H {\sc II} galaxies
may be of great importance if used as a distance indicator of galaxies
at large redshift, since H {\sc II} galaxies are easy to find at great
distances (see also Melnick \etal 2000). Particularly, as the global
line emitting properties reflect the intrinsic properties of the
central core component, observations even with poor spatial resolution
could accurately define the luminosity and $\sigma$ values of the
dominant central core in every galaxy.

\acknowledgments{We thank Roberto Terlevich for a critical reading of
the original version of this paper and for valuable discussion on this
work. This study was partly financed by the Spanish DGES (Direcci\'on
General de Ense\~nanza Superior) (grant PB97-0158). GTT acknowleges
partial support from Conacyt (M\'exico; grant 211290-5-28501E).  GTT
and CMT aknowledge the hospitality of the Observat\'orio Nacional (Rio
- Brasil) where part of this work was carried out.  The WHT is
operated on the island of La Palma by the ING at the Observatorio del
Roque de los Muchachos. Finally, we thank the anonymous referee for his/her
comments that helped  improve the presentation of our results.}


\begin{thebibliography}{}  


\bibitem[Bosch(1999)]{bo} Bosch, G., 1999, Ph.D. thesis, University of
Cambridge

\bibitem[Cair\'os \etal(2000)]{Cai} Cair\'os, L.M., Vilchez, J.M.,
Gonz\'alez--P\'erez, J.N., Iglesias--P\'aramo, J. and Caon, N. 2000,
\apjs, in press

\bibitem[Chu \& Kennicutt(1994)]{chu} Chu, Y-H. \& Kennicutt, R. 1994,
ApJ 425, 720

\bibitem[Doublier \etal(1997)]{doub} Doublier, V., Comte, G.,
Petrosian, A., Surace, C, Turatto, M., 1997, \aandas, 124, 405


\bibitem[Fuentes-Masip \etal(2000)]{Fuent} Fuentes-Masip, O.,
Mu\~noz-Tu\~n\'on,C., Casta\~neda,H.O, \& Tenorio-Tagle, G., 2000,
\aj, 119, 2166



\bibitem[Marlowe \etal(1999)]{marl} Marlowe, A.T., Meurer, G.R. \&
Heckman, T.M., 1999, \apj, 522, 183

\bibitem[Melnick(1979)]{mel} Melnick, J., 1979, \apj, 228, 112

\bibitem[Melnick(1987)]{mel2} Melnick,J., 1987, in {\it \,``Starburst and
Galaxy Evolution\,''}, eds. T.X.Thuan, T.Montmerle \& J.Tran Thanh
Van, editions Fronti\`eres Gif Sur Yvette, France, p. 215

\bibitem[Melnick \etal(1987)]{mmtg} Melnick,J., Moles M., Terlevich
R. \& Garcia-Pelayo J.M., 1987, \mnras, 226, 849

\bibitem[Melnick \etal(1988)]{mtm} Melnick,J., Terlevich R. \& Moles
M., 1988, \mnras, 235, 297

\bibitem[Melnick \etal(2000)]{mtt} Melnick,J., Terlevich,R. \&
Terlevich,E., 2000, \mnras, 311, 629

\bibitem[Mu\~noz-Tu\~n\'on \etal(1994)]{cas} Mu\~noz-Tu\~n\'on, C., 1994,
in Tenorio-Tagle, G. Ed., Violent Star Formation: From 30 Doradus to
QSOs. Cambridge University Press. p 25

\bibitem[Mu\~noz-Tu\~n\'on \etal(1995)]{cas2} Mu\~noz-Tu\~n\'on, C.,
Gavryusev, V. \& Casta\~neda, H., 1995, AJ 110, 1630

\bibitem[Mu\~noz-Tu\~n\'on \etal(1996)]{cas3} Mu\~noz-Tu\~n\'on, C.,
Tenorio-Tagle, G, Casta\~neda, H. \& Terlevich, R., 1996, AJ 112, 1636

\bibitem[Noeske \etal (2000)]{noe} Noeske, K.G., Guseva, N.G., Fricke,
K.J., Izotov, Y.I., Papaderos, P. \& Thuan, T.X., 2000, in ``The
Evolution of Galaxies. I- Observational Clues'', \apspsc, in press

\bibitem[Sabalisck \etal(1995)]{nan} Sabalisck, N., Tenorio-Tagle, G,
Casta\~neda, H., Mu\~noz-Tu\~non, C.  1995, ApJ 444, 200.

\bibitem[Sargent \& Searle(1970)]{ss} Sargent W.L.W. \& Searle L.,
1970, \apjl, 162, L155

\bibitem[Schulte-Ladbeck \etal(1999a)]{shgc} Schulte-Ladbeck, R.E.,
Hopp, U., Greggio, L. \& Crone, M.M., 1999a, \aj, 118, 2705

\bibitem[Schulte-Ladbeck \etal(1999b)]{shcg} Schulte-Ladbeck, R.E., Hopp,
U., Crone, M.M.  \& Greggio, L.,1999b, \apj, 525, 709

\bibitem[Suntzeff \etal(1999)]{spcpga} Suntzeff, N.B., Phillips, M.M.,
Covarrubias, R., Navarrete, M., Peres, J.J.P., Guerra, A., Acevedo,
M.T., Doyle, L.R., Harrison, T., Kane, S., Long, K.S., Maza, J.,
Miller, S., Piatti, A.E., Claria, J.J., Ahumada, A.V., Pritzl, B.,
Winkler, P.F., 1999, \aj, 117, 1175

\bibitem[Taylor(1997)]{} Taylor,C.L 1997, \apj, 480, 524

\bibitem[Telles(1995)]{tel} Telles,E., 1995, Ph.D. thesis, University
of Cambridge

\bibitem[Telles \& Bordalo(2000)]{telbor} Telles,E. \& Bordalo,V.,
2000, in preparation

\bibitem[Telles \& Maddox(2000)]{telmad} Telles,E. \& Maddox,S., 2000,
\mnras, 311, 307

\bibitem[Telles \& Sampson(2000)]{telsam} Telles,E. \& Sampson,L.,
2000, in ``The Evolution of Galaxies. I- Observational Clues'',
\apspsc, in press

\bibitem[Telles \& Terlevich(1993)]{telter} Telles,E. \& Terlevich,R.,
1993, \apspsc, 205, 49

\bibitem[Telles \& Terlevich(1995)]{telter2} Telles,E. \&
Terlevich,R., 1995, \mnras, 275, 1
 
\bibitem[Telles \& Terlevich(1997)]{telter3} Telles,E. \&
Terlevich,R., 1997, \mnras, 286, 183
 
\bibitem[Telles \etal(1997)]{telmt} Telles,E., Melnick,J. \&
Terlevich,R., 1997, \mnras, 288, 78

\bibitem[Telles \etal(1999)]{telttks} Telles,E., Tapia,M.,
Terlevich,R., Kunth,D. \& Sampson,L., 1999, in: K.A. van der Hucht,
G. Koenigsberger \& P.R.J. Eenens (eds.), ``Wolf-Rayet Phenomena in
Massive Stars and Starburst Galaxies'', Proc. IAU Symposium No. 193
(San Francisco: ASP), 622

\bibitem[Telles \etal(2000)]{telttks2} Telles,E., Tapia,M.,
Terlevich,R., Kunth,D. \& Sampson,L., 2000, in preparation

\bibitem[Tenorio-Tagle \etal(1993)]{tmc} Tenorio-Tagle, G.,
Mu\~noz-Tu\~non, C. \& Cox, D., 1993, ApJ 418, 767

\bibitem[Tenorio-Tagle \etal(1996)]{tmcf} Tenorio-Tagle, G.,
Mu\~noz-Tu\~non, C. \& Cid-Fernandes, R., 1996, ApJ 456, 264

\bibitem[Terlevich \& Melnick(1981)]{tm} Terlevich, R. \& Melnick, J.,
1981, \mnras, 195, 839

\bibitem[Terlevich \etal(1991)]{schg} Terlevich, R., Melnick, J.,
Masegosa, J., Moles, M. \& Copetti, M.V.F., 1991, \aandas, 91, 285
(SCHG)

\bibitem[Van Zee \etal(1998)]{vss} Van Zee, L., Skillman, E.D. \&
Salzer, J.J., 1998, \apj, 116, 1186


\end{thebibliography}
\end{document}